# Numerical modeling of distributed combustion without air dilution in a novel ultra-low emission turbulent swirl burner

Dániel Füzesi[1,a)], Milan Malý[2], Jan Jedelský[2], Viktor Józsa[1]

[1]*Department of Energy Engineering, Faculty of Mechanical Engineering, Budapest University of Technology and Economics, Műegyetem rkp. 3., H-1111 Budapest, Hungary*

[2]*Faculty of Mechanical Engineering, Brno University of Technology, Technicka 2896/2, 616 69 Brno, Czech Republic*

**ABSTRACT**

Distributed combustion, often associated with the low-oxygen condition, offers ultra-low $NO_X$ emission. However, it was recently achieved without combustion air dilution or internal flue gas recirculation, using a distinct approach called Mixture Temperature-Controlled combustion. Here, the fuel-air stream is cooled at the inlet to delay ignition and hence foster homogeneous mixture formation. The aim of this numerical study aims to understand the operation of this combustion concept better and present a robust framework for distributed combustion modeling in a parameter range where such operation was not predicted before by any existing theory. Further, liquid fuel combustion was evaluated that brings additional complexity. Four operating conditions were presented at which distributed combustion was observed. The reacting flow was modeled by Flamelet-Generated Manifold, based on a detailed n-dodecane mechanism. The Zimont turbulent flame speed model was used with significantly reduced coefficients to achieve distributed combustion. The droplets of airblast atomization were tracked in a Lagrangian frame. The numerical results were validated by Schlieren images and acoustic spectra. It was concluded that the reactant dilution ratio remained below 0.25 through the combustion chamber, revealing that the homogeneous fuel-air mixture is the principal reason for excellent flame stability and ultra-low $NO_X$ emission without significant internal recirculation. The potential applications of these results are boilers, furnaces, and gas turbines.

[a] Corresponding author. Email: fuzesi.daniel@gpk.bme.hu

# I. INTRODUCTION

Climate summits in the past years made it clear that if humanity wishes to avoid a climate disaster, fossil fuel consumption must be drastically cut back. However, solving the 100 EJ problem[1] needs a magnitude more renewable energy than available today. To mitigate the anthropogenic $CO_2$ emissions without severely compromising our current habits and the economy, advanced fuels and novel technologies are both essential[2]. Gas turbines offer excellent load balancing potential[3], however, fossil fuel-free operation is only expected if carbon emission prices become excessive[4]. Compared to alternative technologies, gas turbines will remain the main power plants in aviation due to their outstanding power density compared to alternative technologies[5]. Flameless or Moderate or Intense, Low-oxygen Dilution (MILD) combustion is highly desired, which offers negligible $NO_X$ emission through distributed combustion[6] and comes with low susceptibility to thermoacoustic instabilities[7]. The present paper aims to provide a robust numerical approach to model an advanced combustion concept that offers nearly distributed combustion without oxygen dilution, i.e., outside the MILD combustion conditions. The results are highly beneficial for all advanced, ultra-low emission combustion chambers.

The rich burn-quick quench-lean burn technology is the least advanced concept on the market, which is otherwise used in most commercial aviation jet engines due to its good flame stability[8]. Lean swirl burners feature notably lower emissions but struggle with thermoacoustic problems at the design point[9]. Catalytic combustors were proposed to mitigate this problem but failed to have a market penetration due to excessive unburnt fuel emission[10]. Porous media burners are under development [11], however, their disadvantage is the low turndown ratio, an essential measure of the current heat engines for both transportation and energy generation.

MILD combustion has been successfully employed in atmospheric industrial combustion systems for decades[12,13], but its application in gas turbines requires further development[7] since neither the extensive exhaust gas recirculation nor inert gas dilution[14] is an option here in industrial scales. Kumar et al.[15] investigated kerosene combustion in a two-stage combustor with tangential air dilution achieving a 75% $NO_x$ drop at distributed combustion mode. Gupta et al.[16] estimated the Damköhler number of colorless distributed combustion in the order of magnitude of $10^{-2}$, concluding that slow reactions characterize this regime. J. A. Wünning and J. G. Wünning[17] determined that exhaust gas recirculation rate of at least two and reactant temperature should exceed the autoignition temperature to achieve flameless combustion.

To eliminate the need for air dilution, the Mixture Temperature-Controlled (MTC) combustion concept was recently introduced[18], delivering distributed combustion and hence ultra-low $NO_x$ emissions by using ambient air as an oxidizer. The difference between MTC and MILD combustion is that the former features no dedicated oxygen dilution method, and the mixture inlet temperature remains below the autoignition temperature. Instead, the mixture is cooled to delay ignition and allow more time for homogeneous mixture formation and reduced volumetric Heat Release Rate (HRR), essential for low emissions. Cooling in the present case was provided by an airblast atomizer that hindered early ignition. Both MTC and MILD combustion can maintain distributed combustion in the sense of reaction zone shape, however, their operation is significantly different since there was no internal or external flue gas recirculation designed according to the criteria map of MILD combustion[7]. For this reason, the extent of flue gas recirculation will also be investigated in this paper, which is a critical measure of similarity.

Following the preliminary Computational Fluid Dynamics (CFD) investigations on diesel and waste cooking oil combustion of the MTC combustion concept[19], it was concluded that unsteady calculations are inevitable to model distributed combustion. The gas-phase was modeled in the Eulerian frame, while the Lagrangian frame was used for liquid fuel droplets of the spray, a standard approach in both diesel engine[20] and jet engine[21] simulations. Reynolds-Averaged Navier-Stokes models can be applied for colorless distributed combustion modeling, focusing on hydrogen[22,23]. Unfortunately, these models are not compatible with the present problem. Highly simplified 2D biodiesel combustion modeling was performed by Dixit et al.[24], using the one-equation eddy dissipation concept (EDC) and steady calculations, which cannot be efficiently implemented in distributed combustion modeling. It was the motivation for developing the generalized EDC[25]. There were other concepts recently developed for MILD combustion, such as Partially Stirred Reactor[26]. As for simpler models, Validi et al.[27] used a complex Large Eddy Simulation (LES) model combined with filtered mass density functions to simulate methane colorless distributed combustion. The LES results were compared to Particle Image Velocimetry (PIV) images, showing a poor match. V. K. Arghode et al.[28] performed steady-state colorless methane combustion simulation, validated by PIV measurements with a mixed agreement. Karyeyen et al.[23] validated their model under non-reacting conditions. Considering the complexity of liquid fuel combustion and the practical geometry, the requirements of neither the EDC nor the Partially Stirred Reactor could not be met in the present case. FGM was currently selected since it is more robust and computationally cheap using look-up tables instead of detailed chemistry calculations[29]. FGM assumes that the flame is an ensemble of one-dimensional flames combined with a manifold approach. Hence, the reaction mechanisms are calculated as one-dimensional laminar flamelets creating manifolds as a function of mean mixture fraction, scalar dissipation, and progress variable[30] to reduce the number of equations and hence the computational time[31].

The novelty of this work is presenting the CFD analysis of distributed combustion of the MTC combustion concept at four setups. Since a swirl burner was used, the results were compared with known characteristics of V-shaped flames and MILD combustion, especially focusing on flue gas recirculation. Validation was performed by comparing the numerical results with Schlieren measurements and acoustic spectra. Since distributed combustion modeling with ambient air as an oxidizer was not published, this paper will provide a guideline for combustion engineers to model distributed combustion without air dilution and develop combustion chambers around it.

## II. MATERIALS AND METHODS

This section starts with a brief introduction of the measurement setup, emphasizing the Schlieren technique for validation. Section II B details the computational mesh and the numerical setup. Lastly, the fuel properties were detailed since the modeling of distributed combustion is sensitive to the material properties, requiring realistic thermophysical data besides considering a detailed reaction mechanism.

### A. Measurement setup

The geometry of the burner is shown in Fig. 1. The airblast atomizer nozzle is blue with a 2.2 mm orifice diameter, and the 45° flat swirl vanes are red with 40 mm tip and 21 mm hub diameters and eight blades. The length of the mixing tube is 100 mm, measured from the atomizer nozzle tip. The yellow part is for gas/auxiliary air injection, which was not used this time. The central pipe for standard diesel fuel (EN590:2017) is grey with 1.2 mm inner and 1.5 mm outer diameter.

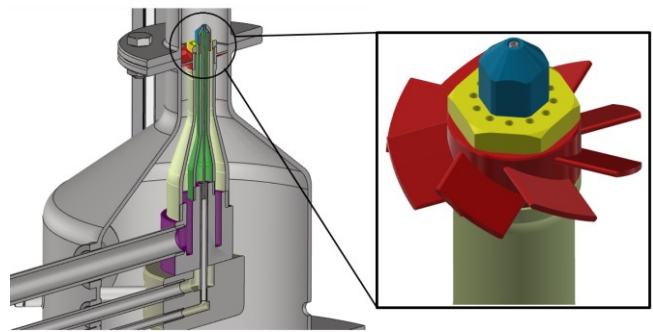

FIG. 1. The geometry of the modeled burner.

Figure 2 shows the schematic measurement setup, including the Schlieren optical system in Z layout. Both the Ø150 mm mirrors had a 750 mm focal length. Image recording was performed at 10 kHz by a FASTCAM SA-Z type 2100K-M-16GB (Photron, Japan) for 0.2 s, acquiring 2,000 images. The flame was illuminated by an HPLS-36DD18B (Lightspeed Technologies, USA) pulsed LED light source triggered by the camera. Combustion noise was recorded for 20 s by a GRAS 146AE microphone at 20 kHz. A 20 dB attenuator was used between the microphone and the pre-amplifier to avoid signal saturation, which was installed during calibration with less than 1 dB deviation from the reference up to 10 kHz, complying with the IEC Standard 61672 class 1. The microphone sensitivity was 5.25 mV/Pa, and the Data Translation DT9837B data acquisition card had 16.1 bits effective and 24 bits total sample size in the -10–10 V input range, meaning 287.4 μV effective and 1.19 μV displayed resolution from oversampling. The cases with equivalence ratio ($\phi$) and atomizing gauge pressure ($p_a$) values are included in Table I. The thermal power was 13.3 kW in all cases. Further details on the measurement setup, including uncertainty of flow, pressure, and temperature sensors, are available in[18].

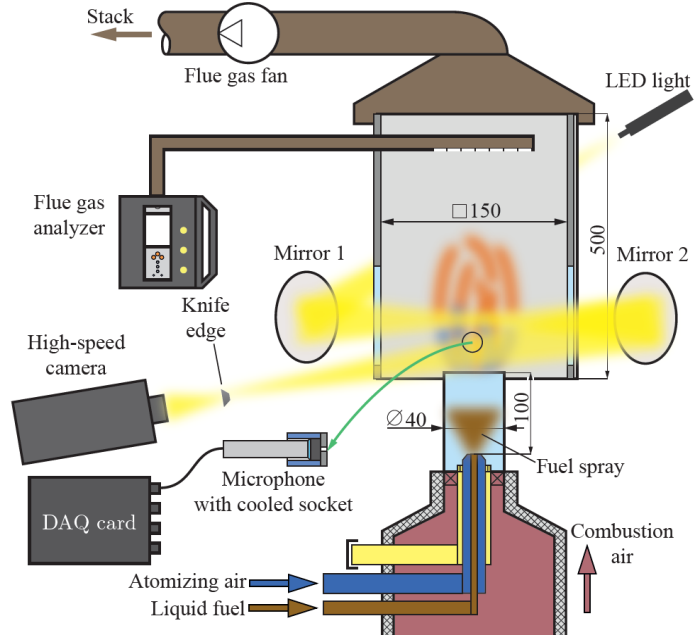

FIG. 2. Schematic of the combustion chamber with the Schlieren setup.

TABLE I. Simulated cases.

| Case №. | 1 | 2 | 3 | 4 |
|---|---|---|---|---|
| $\phi$ | 0.57 | 0.57 | 0.67 | 0.76 |
| $p_a$ | 0.75 | 0.9 | 0.9 | 0.9 |

Uncertainty quantification of velocity calculation from Schlieren images was performed based on the work of Sciacchitano et al. [32], using the published code of the authors, tailoring the calculation parameters to the current case. Table II contains the results at a 95% confidence level. These values are closely a magnitude higher than regular PIV measurements, caused by the fact that the evolution of the tracked density gradients is the consequence of chemical reactions, which are accompanied by thermal expansion. Regular seeding particles are neither deforming nor expanding.

TABLE II. Measurement uncertainty of velocity calculation from Schlieren images at 95% confidence level.

| Uncertainty | Case 1 | Case 2 |
|---|---|---|
| Mean [m/s] | 0.71 | 0.76 |
| Standard deviation [m/s] | 0.2 | 0.19 |
| Maximum [m/s] | 1.54 | 1.68 |

## B. Numerical modeling

The mosaic poly-hexacore mesh of the burner head and the combustion chamber is shown in Fig. 3, considering the real geometry and the best practices[33]. All the contour plots are presented along the y-z plane. A virtual convergent nozzle at the outlet was added to avoid backflow and loss of numerical stability, which is a common solution in combustion chamber simulation[21]. Its geometry was determined by evaluating non-reacting flows and to minimize the effect of the accompanying pressure drop on the flow field. The first half of the combustion chamber was refined since the larger gradients are confined to this region. According to the mesh sensitivity analysis, the final mesh consisted of 318949 cells, considering temperature, velocity magnitude, OH intensity, and droplet diameter distributions. See the supplementary material for further information on mesh sensitivity investigation. The entire numerical problem was solved in the ANSYS Fluent 2021 R1 software environment, including meshing. The pressure-based solver was used in 3D with a time step size of 10 µs, 10 inner iterations per time step, and a total time of 0.2 s for the developed flow. Second-order pressure interpolation scheme, bounded central differencing method for momentum equation, and bounded second-order implicit scheme for the temporal discretization were used. All the other applied numerical schemes and settings can be found in the supplementary material. The convective Courant number was below unity for the majority of the domain, except for the vicinity of the atomizer. Here, its value was peaking at 40, however, the steady air discharge caused no numerical bias, which was checked with a magnitude lower time step size. This large value was caused by the combined effect of the fine mesh, and the two magnitudes larger mean flow velocity than inside the combustion chamber.

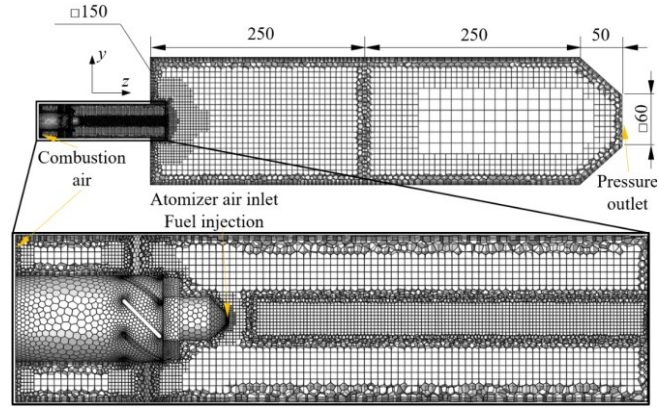

FIG. 3. Cross-section of the control volume (top), and mesh of the burner/mixing tube (bottom).

Spray modeling was included as a steady process to achieve better numerical stability that also comes with a notable computational cost reduction. The Discrete Phase Model of ANSYS Fluent was used for this purpose, using the default set of equations[30]. Consequently, breakup and coalescence were automatically neglected, which is justifiable in diluted sprays[34]. The built-in airblast atomizer model was used to consider the spray, and its half-cone angle was 11°[35] with a fuel temperature of 25 °C. The maximum relative inlet velocity was set according to the corresponding atomizer air inlet pressures [18]. Droplet motion was estimated by the discrete random walk and the random eddy lifetime stochastic models[30].

Diesel fuel was considered as 100% n-dodecane, similar to Ref. 36, using the mechanism of the CRECK Modeling Group, including the thermochemical properties of the species[37]. The chemical reactions were considered by a thermochemical probability density function (PDF)[30], using the premixed FGM model. The 1D flamelet equations were solved in progress variable space, and adiabatic flamelets were considered, while a non-adiabatic PDF table was generated. A similar method was used in Ref. 38 to model MILD combustion, furthermore, the PDF method can be extended to distributed combustion[39] with proper considerations. Combustion was modeled as partially premixed in the present case, using the $C$ Equation[40] since the $G$ Equation is not compatible with FGM. Furthermore, high-fidelity simulation of a real geometry comes with significantly increased computational cost. The PDF table was confined to the 15 most significant species to lower the computational cost of the simulation.

For turbulence-chemistry interaction, the Zimont turbulent flame speed closure model was used, where the turbulent length scale and flame speed constants were set to 0.1, the Schmidt number to 1, and the wall damping coefficient to 0.01. These uncommon parameter values were required to have distributed combustion instead of fast reactions and a straight flame, in line with the model description that explicitly stated that these parameters might need adjustment[30]. The laminar flame speed as a function of the mean mixture fraction was imported from Ref. 41. Thermal radiation was considered by the discrete ordinates model, using the weighted-sum-of-gray-gases model for the gaseous medium[30].

The viscous model was *k-ω* Shear Stress Transport for steady-state, which provided the initial condition for the transient calculations. Here, Scale Adaptive Simulation was used since at least a hybrid viscous model is necessary to properly model the complex flow structures of the presently analyzed semi-industrial burner, generated by the interaction between the swirl vanes and the atomizing free jet. $NO_x$ emission was estimated by the sum of thermal and prompt pathways, calculated from the instantaneous $N_2$, $O_2$, OH, and O concentrations. The boundary conditions (BC) are summarized in Table III.

TABLE III. Boundary conditions.

| BC | Case № | | | |
|---|---|---|---|---|
| | 1 | 2 | 3 | 4 |
| Atomization air inlet [g/s] at 20 °C | 0.754 | 0.831 | 0.831 | 0.831 |
| Combustion air inlet at 200 °C | 7.04 | 6.97 | 5.85 | 5.02 |
| Flue gas pressure outlet gauge pressure [Pa] | 0 | | | |
| Reference pressure [Pa] | 100419 | | | |
| Heat transfer coefficient for the mixing tube wall [W/m²K][42] | 9.77 | | | |
| Heat transfer coefficient for the combustion chamber wall [W/m²K][42] | 8.39 | | | |
| Ambient temperature [°C] | 20 | | | |
| Emissivity[43] | 0.5 | | | |
| Other walls Heat flux [W] | 0 (adiabatic) | | | |
| Fuel inlet [kg/s] | 0.309 | | | |

## C. Fuel properties

Physical material properties of n-dodecane in both liquid and gaseous form were gathered mainly from the National Institute of Standards and Technology (NIST) database[44] and modeled by polynomials. In the case of missing reference data for the evaluated temperature range, estimation methods were used[45]. The required properties of liquid and gas phases and the used methods are listed in Table IV. The boiling point at normal conditions, $T_{b,n}$ of n-dodecane, was 489.3 K, close to 502.5 K, the initial boiling point of the measured diesel fuel sample used for the experiments. Latent heat of evaporation at $T_{b,n}$, $L_{T_{b,n}}$ of $C_{12}H_{26}$ was 256 kJ/kg, and the vapor pressure curve, $p_{vs}$, was determined by the Antoine equation. Liquid-phase density ($\rho_l$), specific heat capacity ($c_{p,l}$), dynamic viscosity ($\mu_l$), and surface tension ($\sigma$) were available from 260 K up to $T_{bn}$ from NIST or by calculation methods. The temperature interval for vapor-phase specific heat capacity ($c_{p,v}$) and mutual diffusion coefficient of vapor and air ($D_{v,a}$) ranged from 280 K to 2000 K, while dynamic viscosity ($\mu_v$), thermal conductivity ($k_v$) of air and products were treated as the temperature-dependent properties of air. The lower heating value of the diesel fuel was 43 MJ/kg, and the stoichiometric air-to-fuel ratio was 14.4 kg air/kg fuel.

TABLE IV. Material properties of n-dodecane, used for modeling diesel fuel.

| | diesel (n-$C_{12}H_{26}$) |
|---|---|
| $T_{bn}$ | NIST |
| $L_{T_{bn}}$ | NIST |
| $\rho_l$ | NIST |
| $c_{p,l}$ | NIST |
| $\mu_l$ | NIST |
| $\sigma$ | Brock[46] |
| $c_{p,v}$ | NIST |
| $\mu_v$ | Lucas[47] |
| $k_v$ | Modified Eucken method[48] |
| $D_{v,a}$ | Fuller[49,50] |
| $p_v$ | NIST |

## III. Results and discussion

The comprehensive evaluation of the CFD results is discussed in this section, which encompasses 0.2 s flow time, presenting both time-averaged and instantaneous data. The first subsection focuses on validating the numerical results by Schlieren images and microphone data. It is followed by the evaluation of temperature and OH distribution, followed by velocity field and analysis in Sections III C, which also includes atomization and turbulence intensity. Since a swirl burner was used, the induced vortex structures are highlighted in Sec. III D. Finally, the $NO_x$ formation was analyzed and compared with measurement data.

### A. Validation

The Schlieren setup was focused on an 86 × 86 mm central field, including the mixing tube outlet. Hence, the presented numerical results in this section are cropped to this area.

Figure 4 shows the mean density gradients, based on 0.2 s real time of 2000 images at 10 kHz sampling frequency, while the considered duration of the transient CFD analysis was also 0.2 s, using 20,000 calculation steps. Unless otherwise stated, these values also apply to the other figures in Section III A. High mean density gradient values are present downstream the lip of the mixing tube on both the measurement and numerical data since cold mixture discharges from the mixing tube, surrounded by flue gas. The mean Schlieren images were normalized. Note that the Schlieren system was not used for analyzing the mixing tube due to the presence of frame bars supporting the combustion chamber and the requirement of very special mirrors to compensate for the glass cylinder. Therefore, only the downstream the lip of the mixing tube is presented here. Both the CFD and the measurement results are asymmetric, hence, the Abel transform was omitted in the latter case. Note that the difference is caused by the planar values of the CFD and the line-of-sight results of the Schlieren technique, which is practically showing the results at the flame front. All cases behave similarly; the magnitude is the lowest in Case 4 since it features the lowest combustion air flow rate and hence the lowest bulk velocity magnitude.

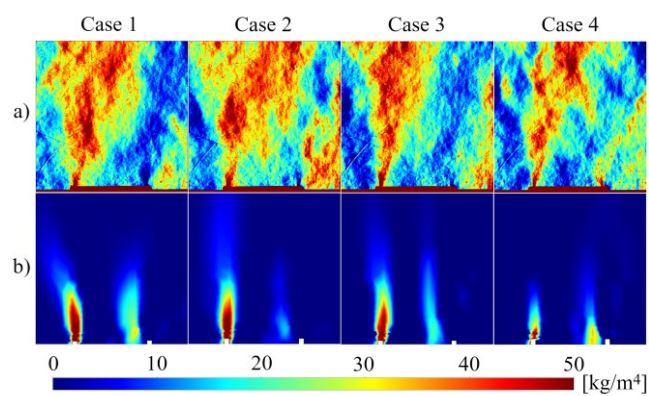

FIG. 4. a) Schlieren mean image and b) mean density gradient of the CFD analysis.

Figure 5 shows the comparison of raw instantaneous Schlieren images and OH distribution of the CFD results. The small, characteristic cellular structures on the Schlieren images show the cold eddies containing the fuel-air mixture, originated from the evaporating droplets. The global boundaries indicate ignition, which was also highlighted by the appearing OH concentration in the CFD analysis. The increased $p_a$ and lower $\phi$ lead to deeper penetration of the cold jet into the combustion chamber. The latter is the combined effect of the increased combustion air flow rate and the decreasing flame speed with $\phi$.

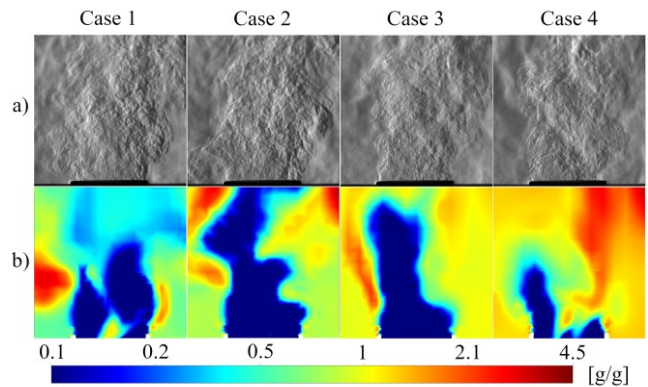

FIG. 5. a) raw instantaneous Schlieren images and b) instantaneous volumetric OH concentration of the CFD analysis. Note the log scale.

The Schlieren images were processed by the PIVLab[51] Matlab package, allowing velocity distribution calculation. Figure 6 presents the measured and computed instantaneous velocity distribution for both cases. The global trends matched, however, the real values are far from each other, most probably due to the line-of-sight measurement of the Schlieren technique, primarily showing the temporal evolution of the reaction zone. The high velocity values near the mixing tube wall highlight that the mixture is forced to the wall, resulting from the swirling flow.

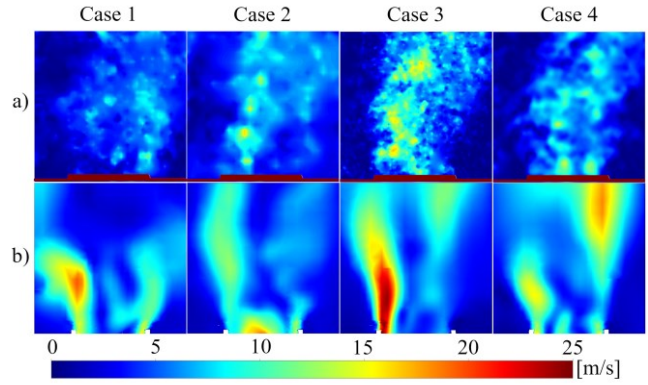

FIG. 6. a) instantaneous velocity profile of Schlieren images and b) the CFD analysis.

The uncertainty of the velocity field calculation, presented above in Table II, is relatively high since instead of particles, the PIV analysis used density gradients in a reacting flow. Here, the fluid packets are, therefore, may deform and expand, unlike the $TiO_2$ seeder particles for combustion measurement.

Heat release determines the temperature field, which governs the local speed of sound. Hence, comparing the mean measured and simulated acoustic spectra is an excellent opportunity for quantitative validation. The time series of the static pressure of the entire 0.2 s transient simulation with 20.0000 samples was used to determine the spectrum, using Fast Fourier Transformation (FFT) with Hamming window, 50% overlap, and 4096 samples. The derived Sound Pressure Level (*SPL*) was compared to the measurement data of Ref. 18, using 20 s measurement data at 20 kHz sampling frequency. The mean *SPL* with identical FFT settings can be seen in Fig. 7. The first and second measured and calculated locally outstanding peaks precisely match. Since the investigated two cases are similar in flame shape and thermal power, there is no notable difference between them since the effect of the atomizing jet well decays before ignition. The first peak at 244 Hz corresponds to the quarter-wave of the combustion chamber, while 844.7 Hz is the one-wave mode of the system from the atomizer nozzle to the combustion chamber outlet.

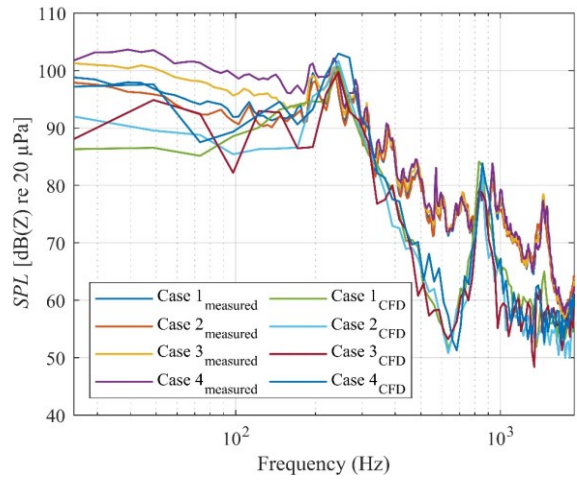

FIG. 7. Measured and numerically calculated spectra.

## B. Combustion characteristics

The time-averaged temperature fields are shown in Fig. 8, and the time evolution is available as a supplementary video for each case. See them as Fig. 8 Case 1 (multimedia view), Fig. 8 Case 2 (multimedia view), Fig. 8 Case 3 (multimedia view), and Fig. 8 Case 4 (multimedia view). The global map for all cases is similar; the difference is caused by the atomizing air jet when comparing Cases 1 and 2, and Cases 2–4 show the effect of combustion air flow rate since $p_a$ and the thermal power are both identical, but $\phi$ varies. The cold air wake of atomizing air discharge is longer with increasing $p_a$ and $\phi$. Due to the increased overall flow rate, Case 2 features the longest cold mixture jet discharge from the mixing tube. It becomes shorter with the increase of $\phi$ through the decreasing combustion air flow rate, and the tip opens to V, similar to Case 1 due to the tangential momentum of combustion air provided by the swirl vanes. Distributed combustion is special a lifted flame, which is visible in all cases. Since the flame front location is determined by the balance of the fresh mixture velocity and the flame propagation speed, the lower combustion air flow rate of Case 4 leads to flame stabilization closed to the mixing tube. Mind the log scale that allows the presentation of both cold air jet discharge and hot flue gas.

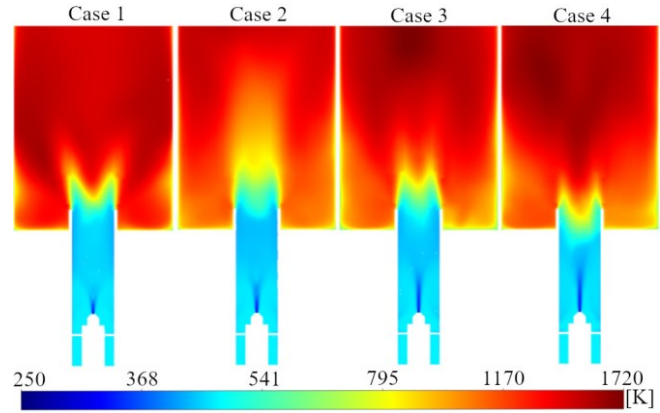

FIG. 8. Mean temperature distribution. Note the log scale. (multimedia view)

The Root Mean Square (RMS) of the temperature is shown in Fig. 9. The mixing tube features notable oscillations arising from the whirling atomizing jet; otherwise, its temperature field is stable, including the hot flue gas 100 mm downstream the mixing tube outlet. According to the simulation, since the axial momentum is smaller in Cases 1 and 4, there is a strong fluctuation right above the mixing tube outlet as the mixture occasionally flashes backward. Due to the increased axial momentum of Cases 2 and 3, the oscillations generally feature lower amplitudes, and they are present at the wake of the mixing tube wall, shown in Fig. 9. Due to the swirling flow, all results follow V-shaped to a varying extent. The Outer Recirculation Zone (ORZ) fluctuations are visible in Case 1, while it is not spectacular in the other three cases.

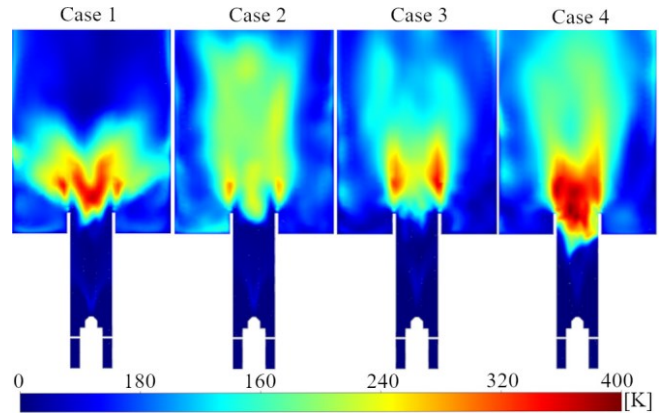

FIG. 9. RMS temperature distribution.

Figure 10 shows the mean OH distribution in a logarithmic scale, representing the heat release that correlates well with the temperature gradient. The temporal evolution of OH is available in videos for Fig. 10 Case 1 (multimedia view), Fig. 10 Case 2 (multimedia view), Fig. 10 Case 3 (multimedia view), and Fig. 10 Case 4 (multimedia view). In all cases, ignition occurs downstream the mixing tube and occupies a large part of the combustion chamber, indicating distributed combustion. The global OH concentration increases principally with $\phi$ since the reactions become more intense as the stoichiometric condition approaches. Hence, the effect of $p_a$ is inferior to $\phi$ since the fuel-air mixture is sufficiently homogeneous in all cases. To visualize the temperature field in 3D, the unsteady mean is presented on five surfaces, shown in Fig. 11. Zero axial distance is the lip of the mixing tube, where the temperature difference is the highest. The cold mixture stream still can be identified 50 mm downstream in Cases 1–3, which wake disappears only at 200 mm for Case 1.

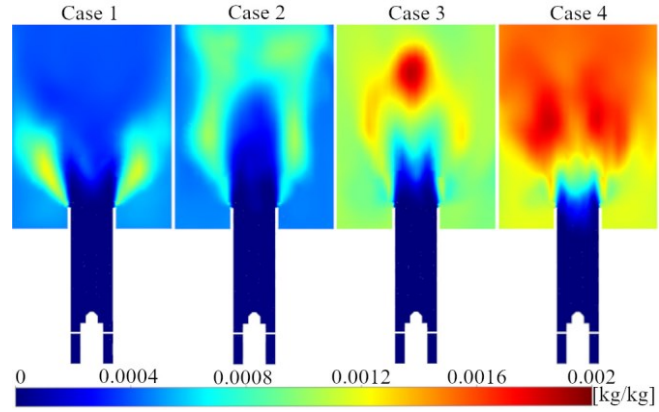

FIG. 10. Mean OH distribution. (multimedia view)

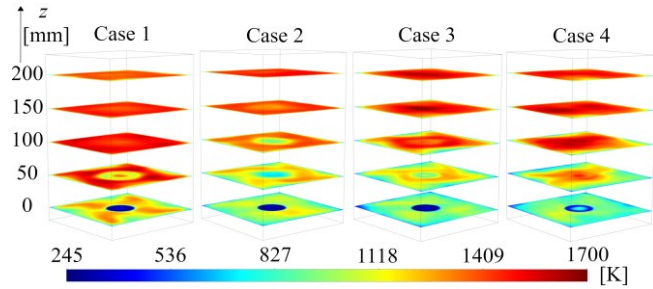

FIG. 11. Mean temperature distribution in five planes, starting from the burner lip.

Figure 12 presents the instantaneous temperature, OH, and $CO_2$ distribution along the x-direction at different downstream distances at $y = 0$. The temperature distribution is the quantitative presentation of that of Fig.8. Hence, higher values are visible in Case 1 with earlier ignition, while closely flat 1500 K is reached from 150 mm in Case 1 and 200 mm in Case 2. Cases 3 and 4 are less even, featuring slight local temperature peaks and a less even profile at $z = 200$ mm. OH represents the heat release, which is more intense in Case 1 and 3 from 25–75 mm. The heat release is notably delayed in Case 2 since it starts at 75 mm and the reactions are not complete even at 200 mm. The OH concentration of Case 4 is significant from 25 mm, and the reaction zone closely ends at 200 mm. $CO_2$, a final product of the reaction, reaches its final value at 100 mm in Case 4, 150 mm in Cases 1 and 3, and 200 mm in Case 2. This implies that the reaction zone is the largest in Case 2 due to low $\phi$ and high $p_g$, and the smallest in Case 4, caused by the low global velocity magnitude due to high $\phi$.

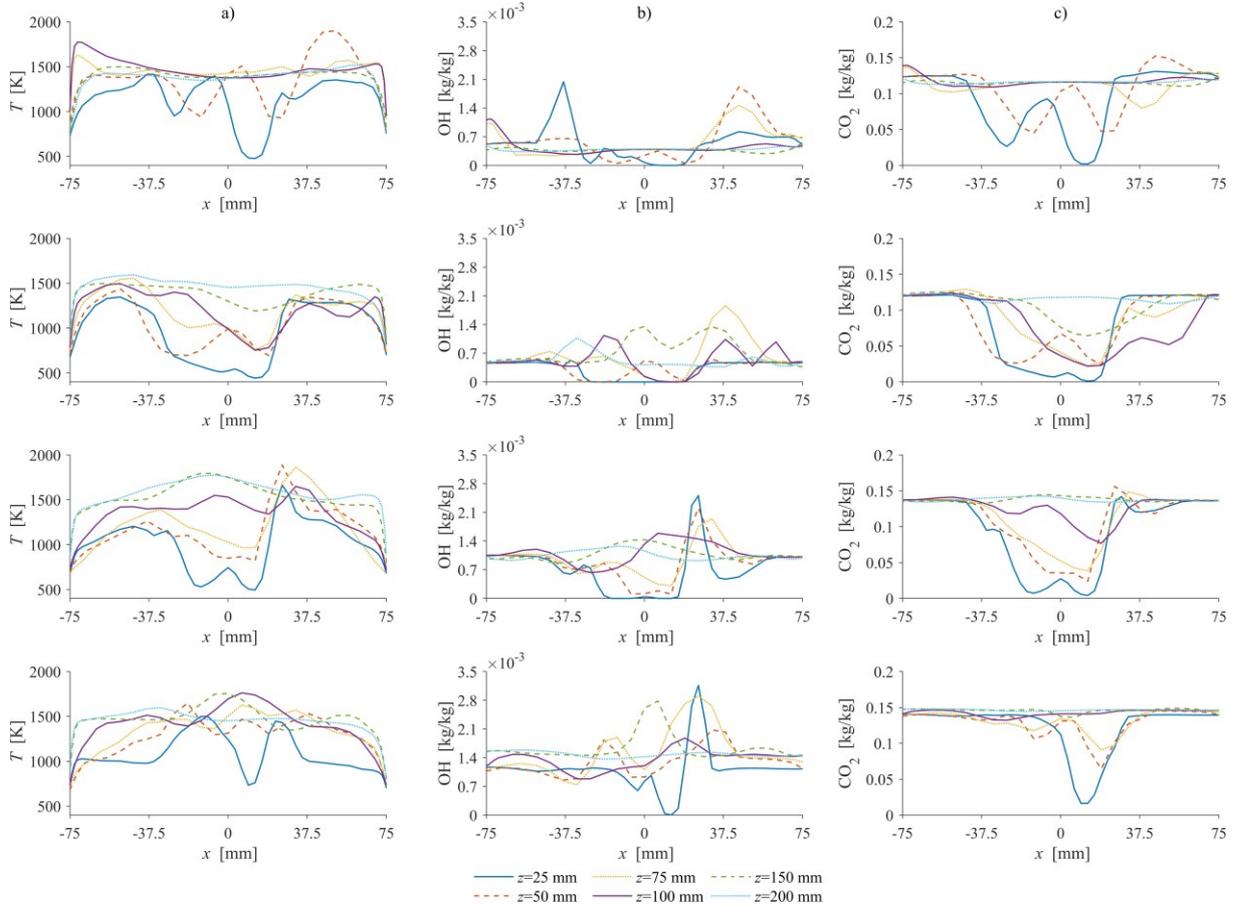

FIG. 12. a) instantaneous temperature, b) OH, and c) $CO_2$ distribution along the x-direction at different axial distances.

## C. Flow field

The temperature field and heat release are key characteristics of a flame, resulting from the two-way coupling between fluid mechanics and chemical reactions. Since a swirl burner is investigated, the swirl number ($S$) is evaluated first. $S$ is defined as[52]:

$$S = \frac{2\pi \int_0^R (Wr)\rho U r \, dr}{2\pi \int_0^R U\rho U r \, dr + 2\pi \int_0^R p r \, dr} \frac{1}{R}, \tag{1}$$

where $R$ is the radius of the mixing tube, $W$ is the tangential velocity, $U$ is the axial velocity, $p$ is the static pressure, $\rho$ is the average density of the mixture, and $r$ is the radial coordinate. $S$ from CFD and the geometric swirl number, $S'$ estimated from the geometry and the boundary conditions[52], are presented in Table V for all cases. The trends are similar, however, there is a 0.08–0.1 offset between the calculated and the estimated values. Its reason is that the static pressure term in Eq. (1) has a notable impact on $S$.

Table V. Calculated and geometric swirl number of all cases.

| Case №. | 1 | 2 | 3 | 4 |
|---|---|---|---|---|
| $S$ | 0.15 | 0.11 | 0.080 | 0.063 |
| $S'$ | 0.25 | 0.22 | 0.17 | 0.14 |

Figure 13 presents the mean velocity magnitude distribution using a log scale. Since $S$ is notably lowered by the high axial momentum of the atomizing jet, Case1 shows only a spectacular V-shaped velocity field, while Cases 2–4 present straight, annular flow. This effect is also present in all the temperature, RMS temperature, and OH fields in Figs. 8–10. The reason for the increased velocity at the top of the images of Cases 3 and 4 is the expansion of the mixture that ignites significantly earlier than that of Case 2, shown earlier in Fig. 8. The velocity decay of the atomizing jets is all confined to the first half of the mixing tube.

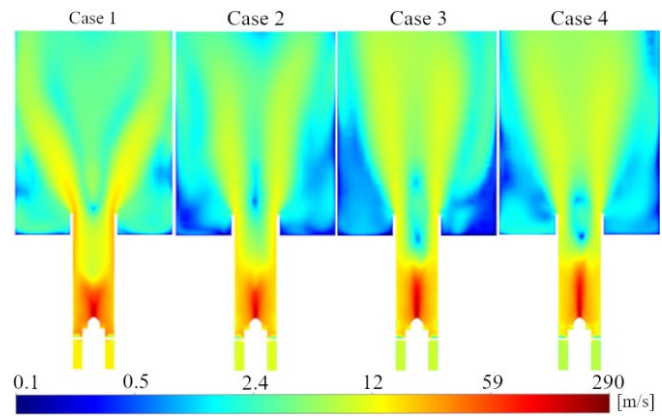

FIG. 13. Average velocity magnitude distribution. Note the log scale.

Figure 14 shows the instantaneous velocity plots. The V-shaped flow field of swirl burners is well-known in the literature[9,53]. The atomizing jet leans to the wall and whirls as in Ref.54; this is why the jet core is short in Case 1 and is different in the structure of Cases 2–4. The swirling flow finally entrains the jet, and the bulk flow is then pushed towards the walls.

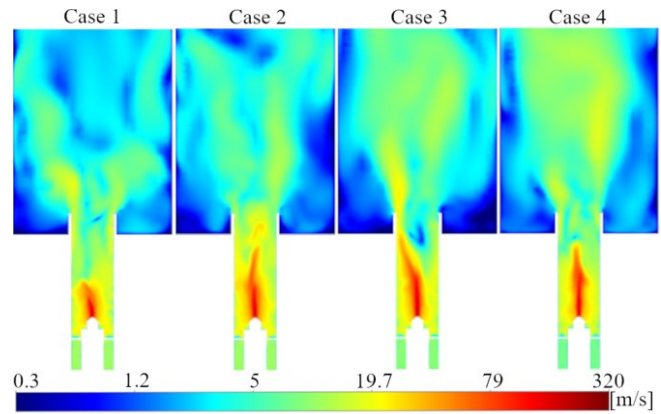

FIG. 14. Instantaneous velocity distribution. Note the log scale.

Instantaneous values are presented in Figure 15 in the x-direction, similar to Fig. 12, which includes velocity data at $y = 0$ and perpendicular to the contour plot of Fig. 14. As the mixture discharges from the chamber, the values decrease and become practically flat from 200 mm in all cases. The wake of the swirl vanes is visible, which matches the shape of the temperature distribution in Fig. 11. After the mixing tube, the mixture takes a V shape and continuously broadens and decays, like free jets in Case 1, Cases 2–4 show annular free jet decay characteristics.

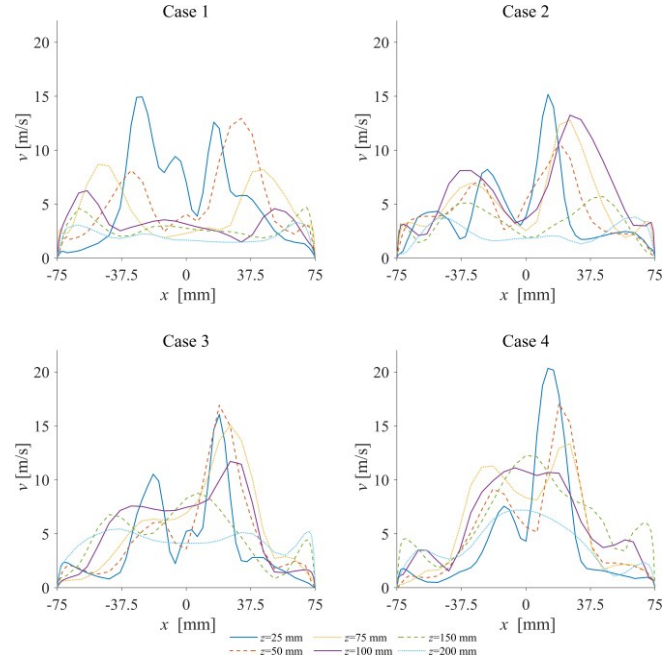

FIG. 15. Instantaneous velocity magnitudes along the x-direction at different axial distances and $y = 0$.

Instantaneous droplet pathways and mean dodecane distribution are presented in Fig. 16. Since droplets are small, they follow the whirling atomizing jet. The numerical results are in line with the observations; no droplet combustion was observed during distributed combustion; all of them were completely evaporated before ignition. The mean evaporated fuel concentration is presented in Fig. 16b, which is the vapor formation from the droplets. The mean is symmetrical, and the highest concentration can be found near the walls as they drift away from the center due to the swirling combustion air. The fuel is present only up to the flame zone, making a perfect match with the low-temperature zone of Fig. 8. The faster evaporation in Cases 2–4 is also attributed to the smaller droplet sizes by the increased $p_a$. Since the atomizing air momentum is smaller in Case 1, the vapor concentration is higher here near the walls. These results are in line with our observations since liquid fuel film forms on the mixing tube wall if its temperature is low due to the lifted flame and the combustion air preheating temperature is insufficient, typical at 150 °C for diesel fuel.

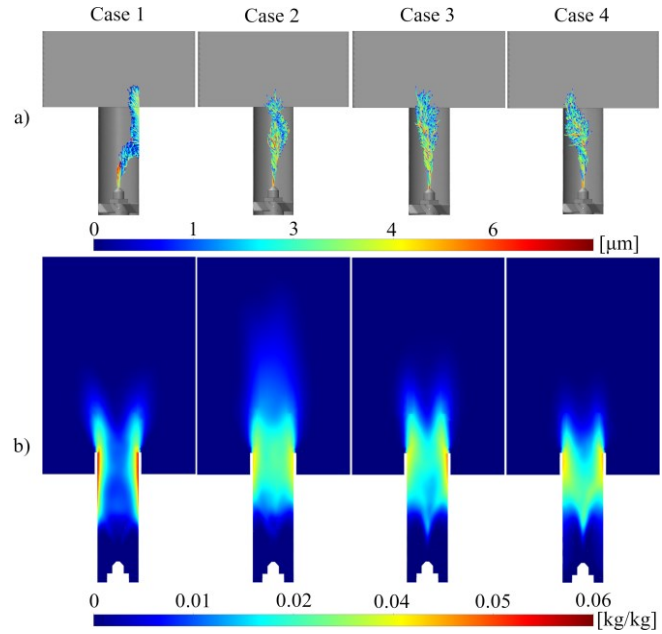

FIG. 16. a) droplet pathways in the Lagrangian frame and b) mean dodecane distribution.

Turbulence intensity describes the velocity fluctuations compared to the mean flow, shown in Fig. 17. Its value is extremely high in the mixing tube due to the two-phase flow, unsteady atomizing jet, and intense swirling flow. Turbulence is also intense in the reaction zone as the gas temperature increases, which amplifies the fluctuations. Finally, it decays in the post-flame zone. The zone volume with extreme is increasing with $p_a$, while that in the combustion chamber shows no general trend with the boundary conditions, and the magnitudes here are similar for all four cases.

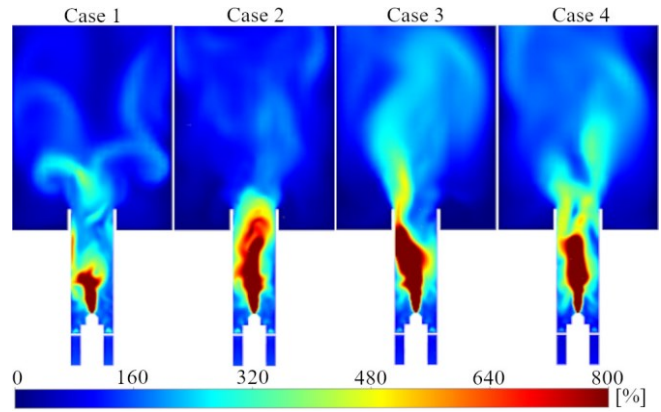

FIG. 17. Instantaneous turbulence intensity distribution.

## D. Vortex structures

Case 1 in Fig. 18 shows the Internal Recirculation Zone (IRZ) by vector plots, colored by the mean OH concentration, which indicated a lifted flame. The ORZ is relatively weak and features a high variation since its structure is poorly localized in the mean vector plot. Downstream of the IRZ, the vector field becomes more even, and uniform flow leaves this combustion chamber region. Since the atomizing air jet momentum is more significant in Cases 2–4, there is no IRZ; instead, a helical flow field is visible. The ORZ occupies about the same volume as the IRZ in Case 1. Even though the outlet of the presented regime is even, the flow field contains several random flow structures due to the irregular vortices downstream of the ORZ.

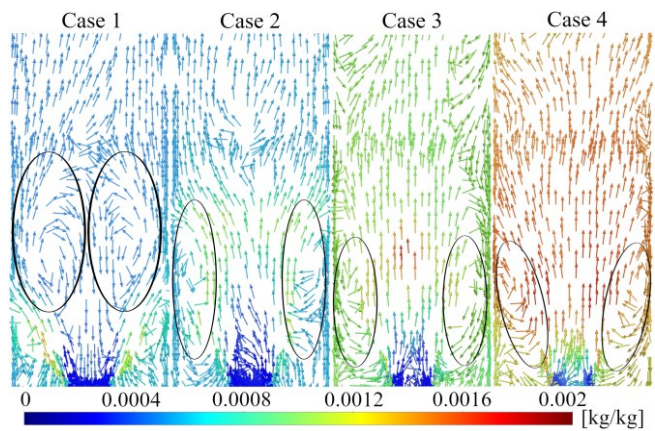

FIG. 18. Mean velocity vector field colored by mean OH concentration.

Figure 19 presents the vector field inside the combustion chamber on four surfaces, colored by the temperature. The swirling motion is clearly visible at all downstream distances due to the conservation of momentum. The high-velocity mixture enters the combustion chamber and entrains the surrounding flue gas, and the entrainment rate increases downstream. The resulting flow global field that is present in the later parts of the combustion chamber as well is nearly developed by reaching 150 mm downstream distance from the mixing tube lip. All distortions correspond to a 3D vortex, as another view of them was shown already in Fig. 18. Since swirl is notably stronger in Case 1, indicated by the V-shaped dominant flow structure, the entire cross-section is occupied by a single large vortex at 150 mm. This is not the case for Cases 2–4 since the high-velocity region stays together at the center. Hence, the flow field is never governed by a single global flow structure; small vortices with random orientation are present. The momentum of the central region decays faster with an increase of $\phi$ since the combustion air mass flow rate decreases.

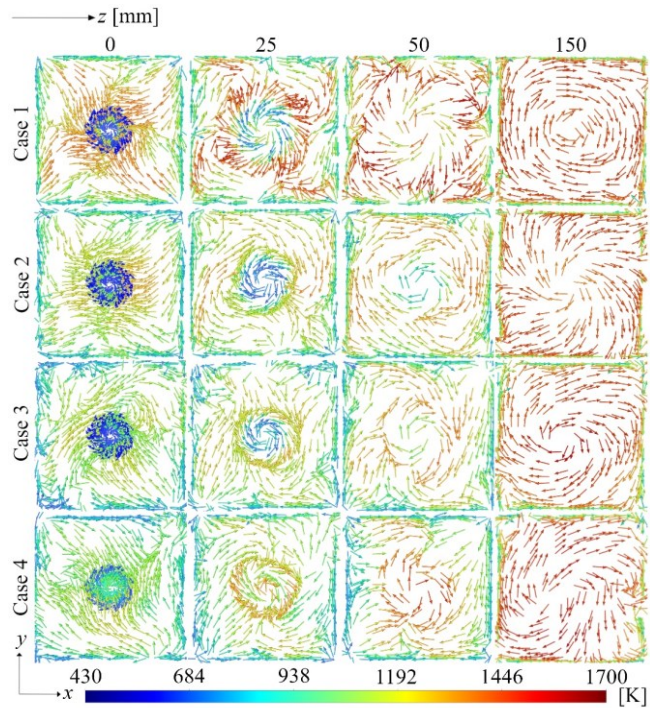

FIG. 19. Mean velocity vector fields at four cross-sections downstream from the burner lip, colored by the unsteady mean temperature.

To visualize the mentioned vortical structures, the $\lambda_2$ criterion was used, and the result is shown in Fig. 20, coloring an iso-surface by temperature. Case 1 clearly shows the coherent IRZ. Cases 2–4 show rather random flow structures, which is an answer to why the blowout stability was excellent for the MTC burner[18]. These vortices facilitate ignition even when the large structures are potentially perturbed as the lean flammability limit is approaching. A precessing vortex core was identified in all cases, however, the higher atomization pressure blew it out in Cases 2–4. Corner vortices are present with lower temperatures and less regular shape than classical ORZ of swirl burners in all cases[53,55]. The randomness of the vortex is also a key to reduced volumetric HRR, which is also present in MILD combustion[56].

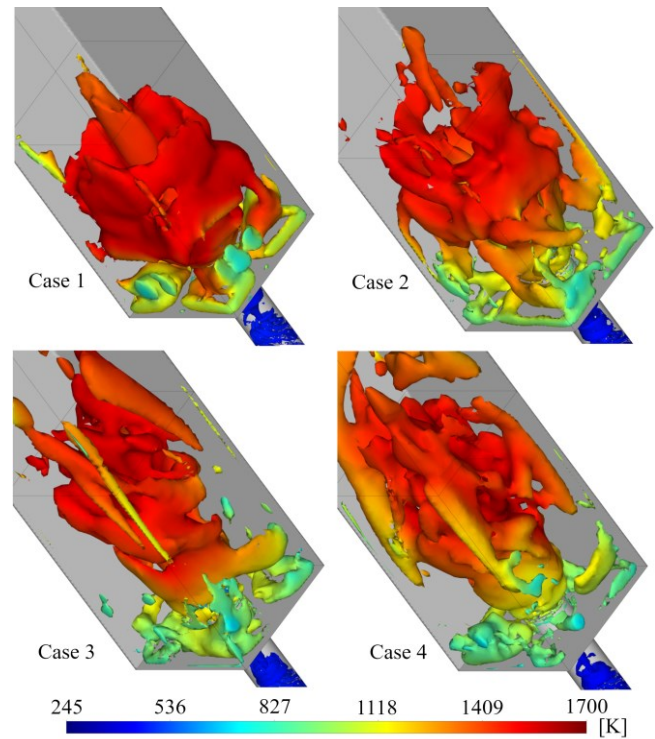

FIG. 20. Mean vortex structures at $\lambda_2 = -3000\ 1/s^2$, colored by temperature.

After seeing the randomness of vortices, the critical question of distributed combustion and the associated unnecessary exhaust gas recirculation or oxidizer dilution by inert gas of the MTC burner is the extent of internal flue gas recirculation since it might lead to mixture dilution and hence low-oxygen condition in the reaction zone, ultimately leading to MILD combustion. There are existing combustion chambers building around this phenomenon[57,58] since intense internal recirculation zones decrease the volumetric HRR. The other possibility of reduced HRR is the lack of internal recirculation, but the flame speed decreases due to the combined effect of the lean mixture and the unique flow structures. To address this question, the reactant dilution ratio, the extent of the recirculation of burnt species to the fresh mixture, can be calculated as[59]:

$$R_{dil} = \frac{|\dot{m}_{ax}| - (\dot{m}_{air} + \dot{m}_{fuel})}{\dot{m}_{air} + \dot{m}_{fuel}}, \qquad (2)$$

where $\dot{m}_{air}$, $\dot{m}_{fuel}$ are the mass flow rate of the total inlet air and the fuel, and $\dot{m}_{ax} = \iint \rho v_{ax} \mathrm{d}x \mathrm{d}y$ is the backflow mass flow rate. The instantaneous results are shown in Fig. 21 along the axial direction of the combustion chamber since the mean was zero in all cases.

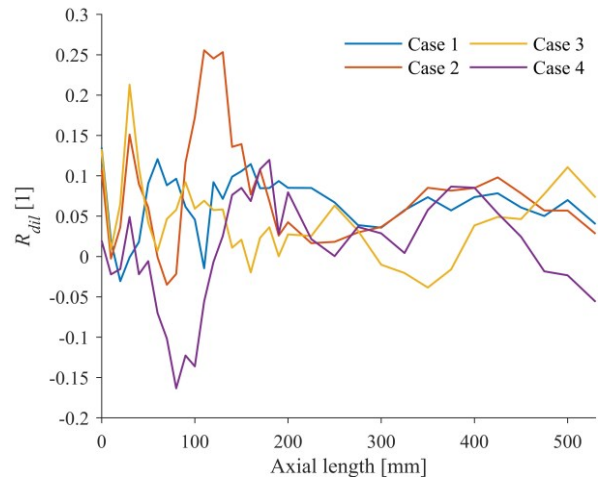

Fig. 21. Reactant dilution ratio in the combustion chamber along the axis. The origin is the mixing tube outlet.

It can be concluded that recirculation is negligible through the present combustion chamber, however, both IRZ and ORZ slightly affect the flow. The intermittent flow behavior arising from evaluating the instantaneous results can explain the negative values. Since the vortical structures, which transfer flue gas back, are not confined to the reaction zone, their presence in the latter part of the combustion chamber is still significant. For comparison, Refs. 59 and 60 show an order of magnitude higher $R_{dil}$ to achieve MILD combustion. In conclusion, the MTC burner achieves distributed combustion principally by cooling the reactants while maintaining a rather random flow field instead of relying on exhaust gas recirculation, like MILD combustion. Consequently, the presented distributed combustion was significantly distinct from MILD combustion and the general conditions provided by Wünning and Wünning[17] for distributed combustion. Currently, no known theory predicts distributed combustion in the presented conditions.

**E. NOx emission**

Figure 22 shows the time-averaged $NO_x$ distribution for all cases. Since the flame temperature was peaking at 1650 K, the thermal $NO_x$ production is already low in the post-flame region. Instead, prompt $NO_x$ production can be identified in the computational domain, which stays low due to the lean conditions. Case 2 is outstanding, probably because this is a very lean case with the highest $p_a$. The combined effect of the two ultimately leads to ultra-low emissions. The mass-weighted average instantaneous values at the combustion chamber outlet and the measured values at 15% $O_2$ are included in Table VI.

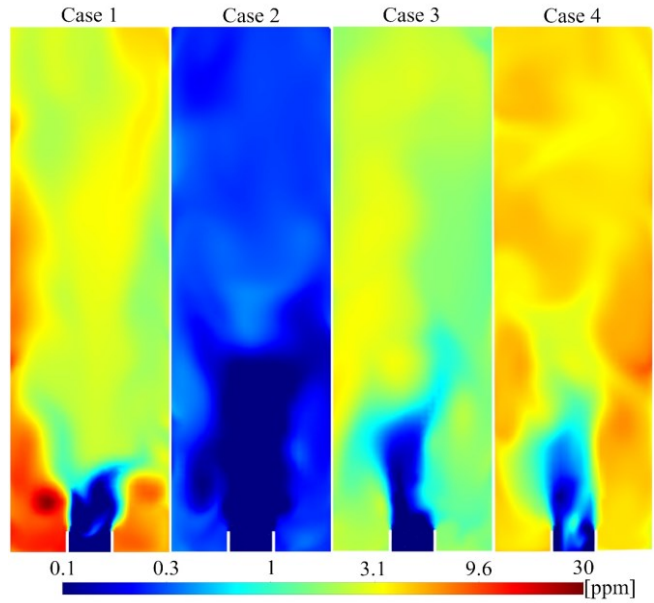

FIG. 22. Instantaneous NO$_x$ distribution at 15% O$_2$ in the dry flue gas. Note the log scale.

Table VI. Measured and simulated NO$_X$ emission in ppm at 15% O$_2$ in the dry flue gas.

| Case №. | 1 | 2 | 3 | 4 |
|---|---|---|---|---|
| NO$_{X,meas}$ | 2.5 | 2 | 3.9 | 4.5 |
| NO$_{X,CFD}$ | 3.2 | 0.3 | 2.2 | 4.4 |

Note that the measurement uncertainty of NO$_X$ emission in this range was 1 ppm, based on calibration with a special gas. Even though the relative deviation is considerable, especially in Case 2, the agreement is good since both the measured and the calculated values are very low in concentration.

CO concentration was measured, and the simulations contain it. However, the deviation here is more than two magnitudes, hence, it is not presented here. The root of this problem is not related to the simplifications used in the present paper since a detailed chemical mechanism in a perfectly stirred reactor simulation also showed similar bias. The reason could be that CO is formed in several magnitudes higher concentrations during combustion than NO$_X$, hence, the reaction parameters should be precisely provided to allow its conversion to CO$_2$, as it is present in real measurements. Consequently, reaction model development could resolve only this discrepancy that is far from the scope of the present paper and requires tremendous effort for real fuels with long-chained hydrocarbons.

## IV. CONCLUSIONS

A numerical investigation of four distributed combustion setups without combustion air dilution was presented in this paper. The conditions were well outside the MILD or flameless combustion zone, already known in the literature, however, distributed combustion could be maintained. The results were validated by Schlieren images, comparing mean density gradient, OH, velocity magnitude, and *SPL*. The agreement between experiments and simulation results is reasonable, implying that the presented framework is robust enough to design practical combustion chambers around distributed combustion without air dilution. The main findings are the following.

The combustion diesel fuel was modeled by FGM with notably reduced turbulent flame speed coefficients in the Zimont model. The necessity of these changes under non-classical premixed flames is known in the literature and falls within model applicability.

The cold discharging mixture from the mixing tube ignited downstream, where complete vaporization of the droplets already occurred. The reaction zone is extended, which characterizes distributed combustion.

The vortical structures differ from classical V-shaped flames and MILD combustion with internal recirculation, making MTC combustion distinct from the known combustion concepts. The reactant dilution ratio was only peaking at 0.25 and remained below 0.1 in the majority of the combustion chamber. The corresponding randomness of the vortices facilitates flame stability and continuous ignition of the upstream fresh mixture.

The ultra-low NO$_X$ formation agrees well with the measurement data. The principal formation pathway is prompt; the contribution of thermal NO is low due to the homogeneous combustion.

## SUPPLEMENTARY MATERIAL

See supplementary material for the mesh sensitivity investigation and numerical schemes.

## ACKNOWLEDGMENTS

The research reported in this paper was supported by the National Research, Development and Innovation Fund of Hungary, project №.s OTKA-FK 124704, 134277, 137758 and TKP2020 NC, Grant No. BME-NC, based on the charter of bolster issued by the NRDI Office under the auspices of the Ministry for Innovation and Technology. The support of the project "Computer Simulations for Effective Low-Emission Energy Engineering" funded as project No. CZ.02.1.01/0.0/0.0/16_026/0008392 by Operational Programme Research, Development and Education, Priority axis 1: Strengthening capacity for high-quality research.

## CONFLICT OF INTEREST

The authors have no conflicts of interest to disclose.

## DATA AVAILABILITY

The data that support the findings of this study are available from the corresponding author upon reasonable request.